\documentclass[12pt,preprint]{emulateapj}

\begin{document}

\title{USco1606-1935: An Unusually Wide Low-Mass Triple System?}

\author{Adam L. Kraus (alk@astro.caltech.edu), Lynne A. Hillenbrand (lah@astro.caltech.edu)}
\affil{California Institute of Technology, Department of Astrophysics, MC 105-24, Pasadena, CA 91125}

\begin{abstract}

We present photometric, astrometric, and spectroscopic observations of 
USco160611.9-193532 AB, a candidate ultrawide ($\sim$1600 AU), low-mass 
($M_{tot}\sim$0.4 $M_{\sun}$) multiple system in the nearby OB 
association Upper Scorpius. We conclude that both components are young, 
comoving members of the association; we also present high-resolution 
observations which show that the primary is itself a close binary system. 
If the Aab and B components are gravitationally bound, the system would 
fall into the small class of young multiple systems which have unusually 
wide separations as compared to field systems of similar mass. However, 
we demonstrate that physical association can not be assumed purely on 
probabilistic grounds for any individual candidate system in this 
separation range. Analysis of the association's two-point correlation 
function shows that there is a significant probability (25\%) that at 
least one pair of low-mass association members will be separated in 
projection by $\la$15\arcsec, so analysis of the wide binary population 
in Upper Sco will require a systematic search for all wide systems; the 
detection of another such pair would represent an excess at the 98\% 
confidence level.

\end{abstract}


\section{Introduction}

The frequency and properties of multiple star systems are important 
diagnostics for placing constraints on star formation processes. This has 
prompted numerous attempts to characterize the properties of nearby 
binary systems in the field. These surveys (e.g. Duquennoy \& Mayor 1991; 
Fischer \& Marcy 1992; Close et al. 2003; Bouy et al. 2003; Burgasser et 
al. 2003) have found that binary frequencies and properties are very 
strongly dependent on mass. Solar-mass stars have high binary frequencies 
($\ga$60\%) and maximum separations of up to $\sim$10$^4$ AU. By 
contrast, M dwarfs have moderately high binary frequencies (30-40\%) and 
few binary companions with separations of more than $\sim$500 AU, while
brown dwarfs have low binary frequencies ($\sim$15\%) and few companions 
with separations $>$20 AU.

The mass-dependent decline in the maximum observed binary separation has 
been described by Reid et al. (2001) and Burgasser et al. (2003) with an 
empirical function which is exponential at high masses 
($a_{max}\propto10^{3.3M_{tot}}$) and quadratic at low masses 
($a_{max}\propto$$M_{tot}^2$). The mechanism that produces the mass 
dependence is currently unknown; N-body simulations show that the 
empirical limit is not a result of dynamical evolution in the field (e.g. 
Burgasser et al. 2003; Weinberg et al. 1987) since the rate of disruptive 
stellar encounters is far too low. This suggests that the limit must be 
set early in stellar lifetimes, either as a result of the binary 
formation process or during early dynamical evolution in relatively 
crowded natal environments. Surveys of nearby young stellar associations 
have identified several unusually wide systems (Chauvin et al. 2004; 
Caballero et al. 2006; ; Jayawardhana \& Ivanov 2006; Luhman et al. 2006, 
2007; Close et al. 2007), but not in sufficient numbers to study their 
properties in a statistically meaningful manner.

We have addressed this problem by using archival 2MASS data to 
systematically search for candidate wide binary systems among all of the 
known members of three nearby young associations (Upper Sco, 
Taurus-Auriga, and Chamaeleon-I; Kraus \& Hillenbrand 2007). Our results 
broadly agree with the standard paradigm; there is a significant deficit 
of wide systems among very low-mass stars and brown dwarfs as compared to 
their more massive brethren. However, we did identify a small number of 
candidate wide systems. One of these candidates is [PBB2002] 
USco160611.9-193532 (hereafter USco1606-1935), a wide (10.87\arcsec; 1600 
AU) pair of stars with similar fluxes and colors. The brighter member of 
the pair was spectroscopically confirmed by Preibisch et al. (2002) to be 
a young M5 star. The fainter member fell just below the flux limit of 
their survey.

In this paper, we describe our photometric, astrometric, and 
spectroscopic followup observations for USco1606-1935 and evaluate the 
probability that the system is an unusually wide, low-mass binary. In 
Section 2, we describe our observations and data analysis methods. In 
Section 3, we use these results to establish that both members of the 
pair are young and co-moving, and that the primary is itself a close 
binary. Finally, in Section 4 we address the possibility that the pair is 
not bound, but a chance alignment of young stars, by analyzing the 
clustering of pre-main-sequence stars in Upper Sco.

\section{Observations and Data Analysis}

Most binary surveys, including our discovery survey, identify companions 
based on their proximity to the primary star and argue for physical 
association based on the (usually very low) probability that an unbound 
star would have been observed in chance alignment. However, the 
probability of contamination is much higher for very wide systems like 
USco1606-1935, so we decided to pursue additional information in order to 
confirm its multiplicity and further characterize its system components. 
In this section, we describe our followup efforts: a search of publicly 
available databases to obtain additional photometry and astrometry, 
acquisition of intermediate-resolution spectra to measure the secondary 
spectral type and test for signatures of youth, and acquisition of 
high-resolution images to determine if either component is itself a 
tighter binary and to test for common proper motion.

\subsection{Archival Data}

We identified USco1606-1935 AB as a candidate binary system using 
archival data from 2MASS (Skrutskie et al. 2006). The binary components 
are bright and clearly resolved, so we were able to retrieve additional 
photometry and astrometry from several other wide-field imaging surveys. 
We collated results for the binary components themselves and for nearby 
field stars from 2MASS, the Deep Near Infrared Survey (DENIS; Epchtein et 
al. 1999), United States Naval Observatory B1.0 survey (USNO-B; Monet et 
al. 2003), and the SuperCOSMOS Sky Survey (SSS; Hambly et al. 2001). The 
DENIS and 2MASS source catalogues are based on wide-field imaging surveys 
conducted in the optical/NIR ($IJK$ and $JHK$, respectively) using 
infrared array detectors, while the USNO-B and SSS source catalogues are 
based on independent digitizations of photographic plates from the First 
Palomar Observatory Sky Survey and the ESO Southern-Sky Survey.

\subsubsection{Photometry}

After evaluating the data, we decided to base our analysis on the $JHK$ 
magnitudes measured by 2MASS and the photographic $I$ magnitude of USNO-B 
(hereafter denoted $I2$, following the nomenclature of the USNO-B 
catalog, to distinguish it from Cousins $I_C$). We chose these 
observations because their accuracy can be directly tested using the 
independent $IJK$ magnitudes measured by DENIS; this comparison shows 
that the fluxes are consistent within the uncertainties. We do not 
directly use the DENIS observations because they are not as deep as the 
other surveys. We adopted the photometric uncertainties suggested in each 
survey's technical reference.

\subsubsection{Astrometry}

As we describe in Section 3.3, there appear to be large systematic 
differences in the astrometry reported by the USNO-B and SSS source 
catalogs. These surveys represent digitizations of the same photographic 
plates, so these systematic discrepancies suggest that at least one survey 
introduces systematic biases in the digitization and calibration process. 
Given the uncertainty in which measurements to trust, we have chosen to 
disregard all available photographic astrometry and only use results from 
2MASS and DENIS. 

Our discovery survey already measured 2MASS relative astrometry for each 
filter directly from the processed atlas images, so we have adopted those 
values. We extracted DENIS astrometry from the source catalog, which 
contains the average positions for all three filters. Both surveys quote 
astrometric uncertainties of 70-100 mas for stars in the brightness range 
of our targets, but that value includes a significant systematic term 
resulting from the transformation to an all-sky reference frame. We have 
conducted tests with standard binary systems of known separation which 
suggest that relative astrometry on angular scales of $<$1\arcmin\, is 
accurate to $\sim$40 mas, so we adopt this value as the astrometric 
uncertainty for each survey.

\subsection{Optical Spectroscopy}

We obtained an intermediate-resolution spectrum of USco1606-1935 B with the 
Double Spectrograph (Oke \& Gunn 1982) on the Hale 5m telescope at Palomar 
Observatory. The spectrum presented here was obtained with the red channel 
using a 316 l/mm grating and a 2.0\arcsec\, slit, yielding a spectral 
resolution of $R\sim$1250 over a wavelength range of 6400-8800 angstroms. 
Wavelength calibration was achieved by observing a standard lamp after the 
science target, and flux normalization was achieved by observation of the 
spectrophotometric standard star Feige 34 (Massey et al. 1988). The spectrum 
was processed using standard IRAF\footnote{IRAF is distributed by the National 
Optical Astronomy Observatories, which are operated by the Association of 
Universities for Research in Astronomy, Inc., under cooperative agreement with 
the National Science Foundation.} tasks.

Our field and young spectral type standards were drawn from membership 
surveys of Upper Sco and Taurus by Slesnick et al. (2006a, 2006b) which 
used identical instrument settings for the spectroscopic confirmation of 
photometrically selected candidate members.

\subsection{High-Resolution Imaging}

We observed USco1606-1935 A and B on February 7, 2006 (JD=2453773) using 
laser guide star adaptive optics (LGSAO; Wizinowich et al. 2006) on the 
Keck-II telescope with NIRC2 (K. Matthews, in prep), a high spatial 
resolution near-infrared camera. The seeing was average to poor 
($\ga$1\arcsec) for most of the observing run, but the system delivered 
nearly diffraction-limited correction in $K'$ (60 mas FWHM) during the 
period of these observations. The system performance was above average 
given the low elevation (34 degrees; 1.8 airmasses), most likely due to 
the proximity and brightness of the tip-tilt reference star ($R=14.2$, 
$d=14\arcsec$).

Images were obtained using the $K'$ filter in both the narrow and wide 
camera modes. The pixel scales in these modes are 9.942 mas pix$^{-1}$ 
(FOV=10.18\arcsec) and 39.686 mas pix$^{-1}$ (FOV=40.64\arcsec). All 
wide-camera observations were centered on the close Aab binary. The A and 
B components were too wide to fit reasonably into a single narrow-camera 
exposure, so we took separate exposure sequences centered on each. We 
obtained four wide-camera exposures of the AB system, seven narrow-camera 
exposures of A, and four narrow-camera exposures of B; the total 
integration times for each image set are 80s, 175s, and 100s, 
respectively. Each set was produced with a 3-point box dither pattern 
that omitted the bottom-left position due to higher read-noise for the 
detector in that quadrant. Single exposures were also taken at the 
central position.

Our science targets are relatively bright, so all observations were taken 
in correlated double-sampling mode, for which the array read noise is 38 
electrons/read. The read noise is the dominant noise term for identifying 
faint sources, yielding $10\sigma$ detection limits of $K\sim19.2$ for the 
wide camera observations, $K\sim18.8$ for the narrow-camera observations 
centered on component A, and $K\sim18.3$ for the narrow-camera 
observations centered on component B; the detection limits for B are 
slightly shallower due to the shorter total integration time. The data 
were flat-fielded and dark- and bias-subtracted using standard IRAF 
procedures. The images were distortion-corrected using new high-order 
distortion solutions (P. Cameron, in prep) that deliver a significant 
performance increase as compared to the solutions presented in the NIRC2 
pre-ship 
manual\footnote{http://www2.keck.hawaii.edu/realpublic/inst/nirc2/}; the 
typical residuals are $\sim$4 mas in wide camera mode and $\sim$0.6 mas in 
narrow camera mode. We adopt these systematic limits as the uncertainty in 
astrometry for bright objects; all faint objects ($K\sim$16-18) have 
larger uncertainties ($\sim$10 mas) due to photon statistics.

We measured PSF-fitting photometry and astrometry for our sources using 
the IRAF package DAOPHOT (Stetson 1987), and specifically with the ALLSTAR 
routine. We analyzed each frame separately in order to estimate the 
uncertainty in individual measurements and to allow for the potential 
rejection of frames with inferior AO correction; our final results 
represent the mean value for all observations in a filter.

In the wide-camera observations, we produced a template PSF based on the 
B component and the field star F1 (see Section 3.1 and Figure 1), both of 
which appear to be single sources. In the narrow-camera observations 
centered on A or B, the science target was the only bright object 
detected in our observations, so there was not a separate source from 
which to adopt a template PSF. We could have adopted a template PSF from 
another set of observations, but the AO correction usually varies 
significantly between targets since it is very sensitive to the seeing, 
elevation, laser return, and tip-tilt separation and brightness. We found 
that no other target in our survey provided a good PSF match.

We addressed this issue for the Aab binary pair by developing a procedure 
to reconstruct the single-source PSF directly from the observations of the 
binary system. Our algorithm begins with a preliminary estimate of the 
single-source PSF, then iteratively fits both components of the binary 
system with the estimated PSF and uses the synthetic PSF to subtract the 
best-fit estimate of the secondary flux. This residual image (which is 
dominated by the primary flux distribution) is then used to fit an 
improved estimate of the single-source PSF.
 
DAOPHOT characterizes an empirical PSF in terms of an analytical function 
and a lookup table of residuals, so we first iterated the procedure using 
a purely analytical function until it converged, then added a lookup 
table to the estimated PSF and iterated until its contents also 
converged. Observations of single stars suggested that the penny2 
function (a gaussian core with lorentzian wings) would provide the best 
analytic fit, so we chose it as our analytic function. Four iterations of 
the fitting process were required for the analytic function to converge 
and 3 iterations were required for the lookup table to converge. Our 
algorithm does not work for the B component because it appears to be 
single, so we adopted the average synthetic single-source PSF from 
analysis of the Aab system to perform PSF fitting and verify that it is 
single.

We calibrated our photometry using 2MASS $K$ magnitudes for the A and B 
components and the nearby field star F1 (Section 3). The 2MASS 
observations were conducted using the $K_s$ filter rather than $K'$, but 
the theoretical isochrones computed by Kim et al. (2005) for the $K_s$ 
and $K'$ systems differ by $\la$0.01 magnitudes for objects in this color 
range; this is much smaller than other uncertainties in the calibration. 
Carpenter (2001) found typical zero point shifts of $\la$0.03 magnitudes 
between 2MASS $K_s$ and several standard $K$ bandpasses, all of which are 
more distinctly different from $K_s$ than $K'$, which also demonstrates 
that the zero point shift between $K_s$ and $K'$ should be negligible.

The calibration process could introduce systematic uncertainties if any 
of the three calibration sources are variable, but based on the small 
deviation in the individual calibration offsets for each source (0.03 
mag), variability does not appear to be a significant factor. We tested 
the calibration using DENIS $K$ magnitudes and found that the two methods 
agree to within 0.01 mag, albeit with a higher standard deviation (0.12 
mag) for DENIS.

\section{Results}

\begin{figure*}
\epsscale{1.00}
\fbox{\plotone{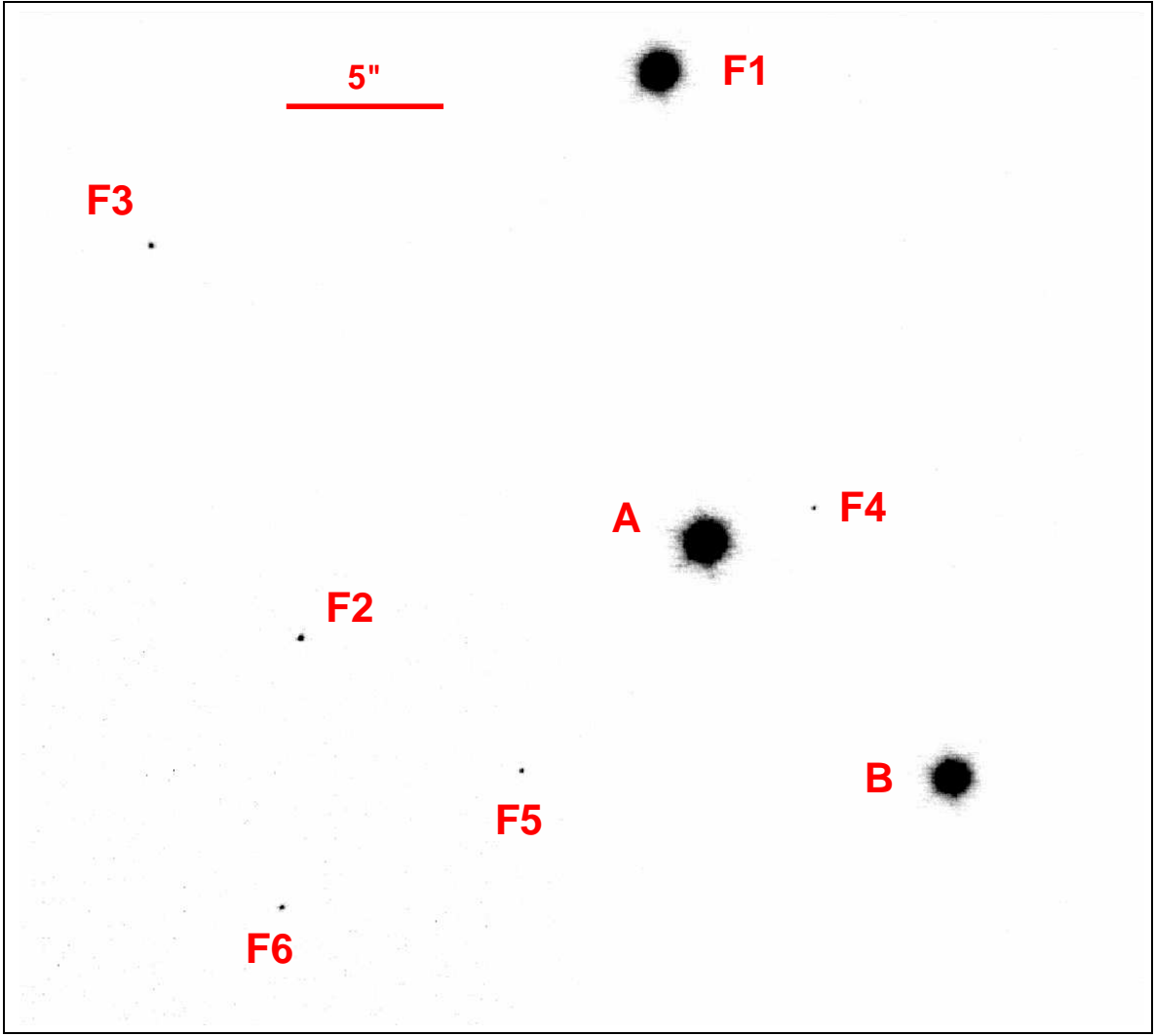}}
\caption{The field surrounding USco1606-1935. The A and B components are 
labeled, as are 6 apparent field stars. The separation between the Aa and 
Ab components is too small to be apparent in this image.} 
\end{figure*}

\begin{figure*}
\epsscale{1.00}
\plotone{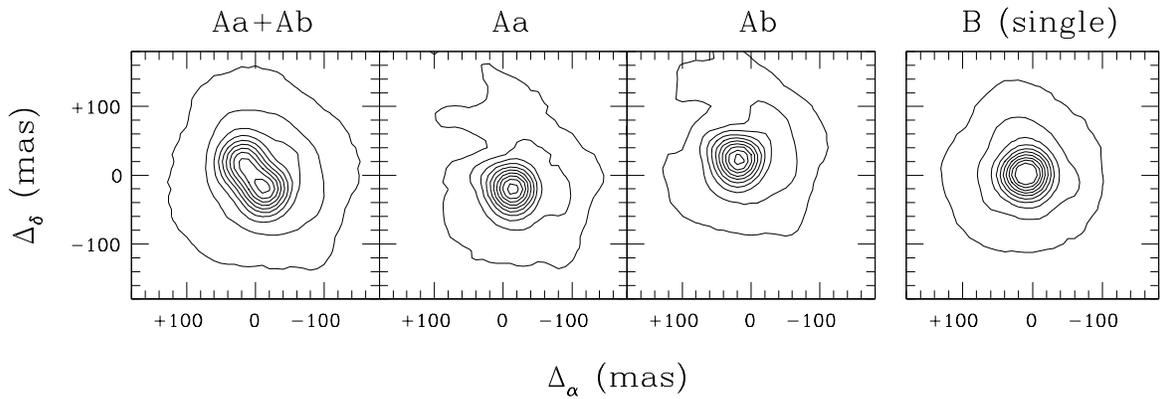}
\caption{Contour plots showing our LGSAO observations of USco1606-1935. The 
first panel shows an original exposure for the Aab pair, the second and third 
panels show Aa and Ab after subtracting best-fit values for the other 
component, and the last panel shows an original exposure for B. The contours 
are drawn at 5\% to 95\% of the peak pixel values.}
\end{figure*}

\begin{figure*}
\epsscale{1.00}
\plotone{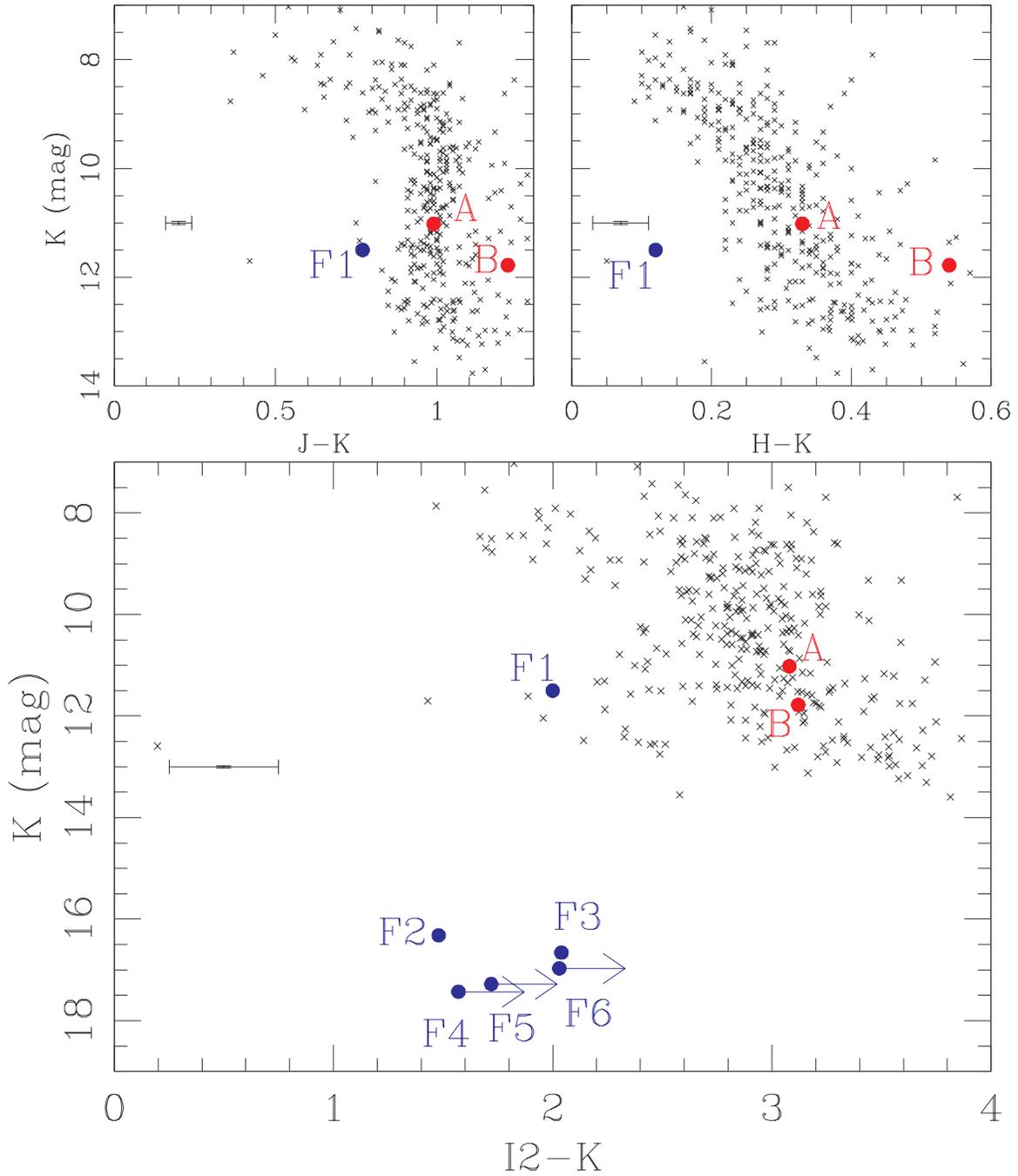}
\caption{Color-magnitude diagrams showing all spectroscopically-confirmed 
members of Upper Sco (black crosses), the A and B binary components (red), and 
the other six objects detected in our LGSAO images (blue). The NIR CMDs (top) 
demonstrate that F1 lies significantly below the association sequence, and 
therefore is an unrelated field star. The optical-NIR CMD (bottom) supports 
this identification and demonstrates that F2 and F3 are also field stars that 
lie below the association sequence. We measure formal upper limits only for 
stars F4-F6, but marginal $R$ band detections in the POSS plates suggest that 
F4 and F6 are also field stars. Typical uncertainties are plotted on the 
left edge of each plot.}
\end{figure*}

\begin{deluxetable*}{lllccccc}
\tabletypesize{\scriptsize}
\tablewidth{0pt}
\tablecaption{Coordinates and Photometry}
\tablehead{\colhead{Name} & \colhead{RA\tablenotemark{a}} & 
\colhead{DEC\tablenotemark{a}} &
\colhead{$K_{LGS}$\tablenotemark{b}} & 
\colhead{$K_{2MASS}$\tablenotemark{b}} & 
\colhead{$H$\tablenotemark{b}} & 
\colhead{$J$\tablenotemark{b}} & 
\colhead{$I2$\tablenotemark{b}}
}
\startdata
A&16 06 11.99&-19 35 33.1&11.04&11.02&11.35&12.01&14.1\\
Aa&-&-&11.71&-&-&-&-\\
Ab&-&-&11.88&-&-&-&-\\
B&16 06 11.44&-19 35 40.5&11.74&11.78&12.32&13.00&14.9\\
F1&16 06 12.09&-19 35 18.3&11.51&11.50&11.62&12.27&13.5\\
F2&16 06 12.90&-19 35 36.1&16.32&-&-&-&17.8\\
F3&16 06 13.23&-19 35 23.7&16.66&-&-&-&18.7\\
F4&16 06 11.75&-19 35 32.0&17.43&-&-&-&-\\
F5&16 06 12.40&-19 35 40.3&17.28&-&-&-&-\\
F6&16 06 12.94&-19 35 44.6&16.97&-&-&-&-\\
\enddata
\tablecomments{Photometry is drawn from our observations ($K_{LGS}$), 2MASS 
($JHK_{2MASS}$), and the USNO-B1.0 catalogue ($I2$). }
\tablenotetext{a}{Coordinates are derived from the 2MASS position for 
USco1606-1935 A and the relative separations we measure using LGSAO. The 
absolute uncertainty in the 2MASS position with respect to the 
International Coordinate Reference System (ICRS) is $\la$0.1\arcsec.}
\tablenotetext{b}{Photometric uncertainties are $\sim$0.03 mag for LGSAO 
and 2MASS photometry and $\sim$0.25 mag for USNO-B1.0 photometry.}
\end{deluxetable*}

\begin{deluxetable*}{lllcccccccccccccc}
\tabletypesize{\scriptsize}
\tablewidth{0pt}
\tablecaption{Relative Astrometry}
\tablehead{\colhead{} & \multicolumn{2}{c}{LGSAO $K$} &
\multicolumn{2}{c}{2MASS $K$} & \multicolumn{2}{c}{2MASS $H$} & 
\multicolumn{2}{c}{2MASS $J$} & \multicolumn{2}{c}{DENIS $IJK$}
\\
\colhead{} & 
\multicolumn{2}{c}{(JD=2453773)} & 
\multicolumn{2}{c}{(JD=2451297)} &
\multicolumn{2}{c}{(JD=2451297)} & 
\multicolumn{2}{c}{(JD=2451297)} &
\multicolumn{2}{c}{(JD=2451332)}
\\
\colhead{} 
& \colhead{$\Delta_{\alpha}$} & \colhead{$\Delta_{\delta}$}
& \colhead{$\Delta_{\alpha}$} & \colhead{$\Delta_{\delta}$}
& \colhead{$\Delta_{\alpha}$} & \colhead{$\Delta_{\delta}$}
& \colhead{$\Delta_{\alpha}$} & \colhead{$\Delta_{\delta}$}
& \colhead{$\Delta_{\alpha}$} & \colhead{$\Delta_{\delta}$}
}
\startdata
Aa&-0.0132&-0.0149&-&-&-&-&-&-&-&-\\
Ab&+0.0201&+0.0266&-&-&-&-&-&-&-&-\\
B&-7.825&-7.460&-7.757&-7.455&-7.749&-7.395&-7.834&-7.382&-7.865&-7.448\\
F1&+1.453&+14.844&+1.401&+14.762&+1.446&+14.732&+1.479&+14.735&+1.418&+14.728\\
F2&+12.839&-3.017&-&-&-&-&-&-&-\tablenotemark{a}&-\tablenotemark{a}\\
F3&+17.571&+9.370&-&-&-&-&-&-&-&-\\
F4&-3.438&+1.056&-&-&-&-&-&-&-&-\\
F5&+5.805&-7.224&-&-&-&-&-&-&-&-\\
F6&+13.385&-11.540&-&-&-&-&-&-&-&-\\
\enddata
\tablecomments{The zero-point for all coordinate offsets is the photocenter of 
the unresolved Aab system. The relative astrometric uncertainties for 
2MASS and DENIS results are $\sim$40 mas; uncertainties for the LGSAO 
results are $\sim$5 mas for bright objects and $\sim$10 mas for faint 
objects.}
\tablenotetext{a}{F2 was marginally detected in $i$ by DENIS, but the 
astrometry is not sufficiently precise to be useful in calculating its proper 
motion.}
\end{deluxetable*}

\begin{figure*}
\epsscale{1.00}
\plotone{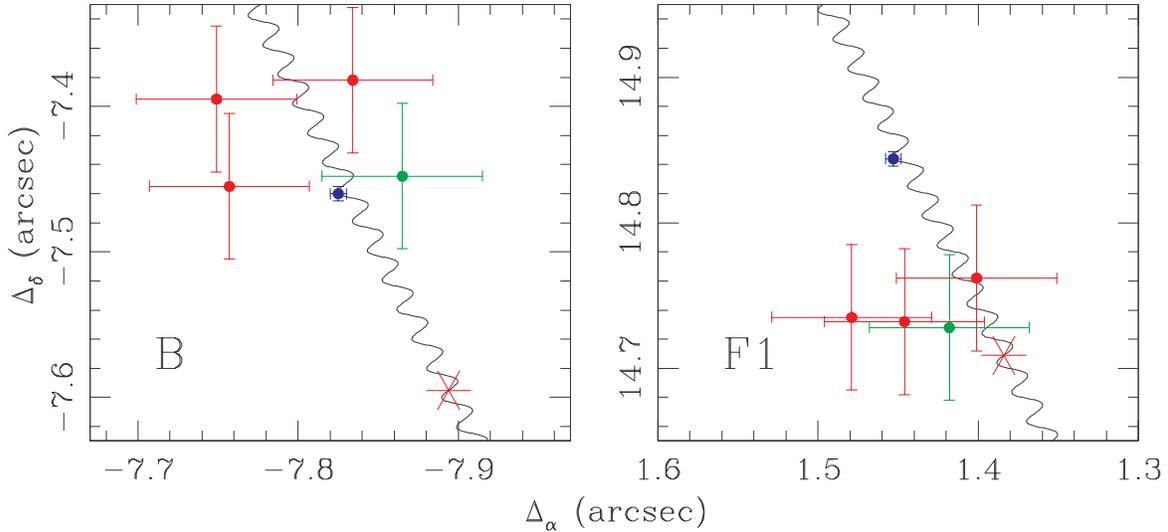}
\caption{Relative separations from the A component to the B component (left) 
and the field star F1 (right) for our LGSAO data and archival 2MASS/DENIS 
data. The blue circles denote LGSAO data, the red circles denote 2MASS data 
for each filter ($J$, $H$, and $K$), and the green circles denote the 
average DENIS values for all three filters ($IJK$). The black line shows the 
expected relative astrometry as a function of time for a stationary object, 
and the predicted archival astrometry values for the non-moving 
(background) case are shown on these curves with red asterisks. The 
results for component B are consistent with common proper motion; the 
results for F1 are inconsistent with common proper motion and suggest 
that the total proper motion is small, denoting a probable background 
star.}
\end{figure*}

\begin{figure}
\epsscale{1.00}
\plotone{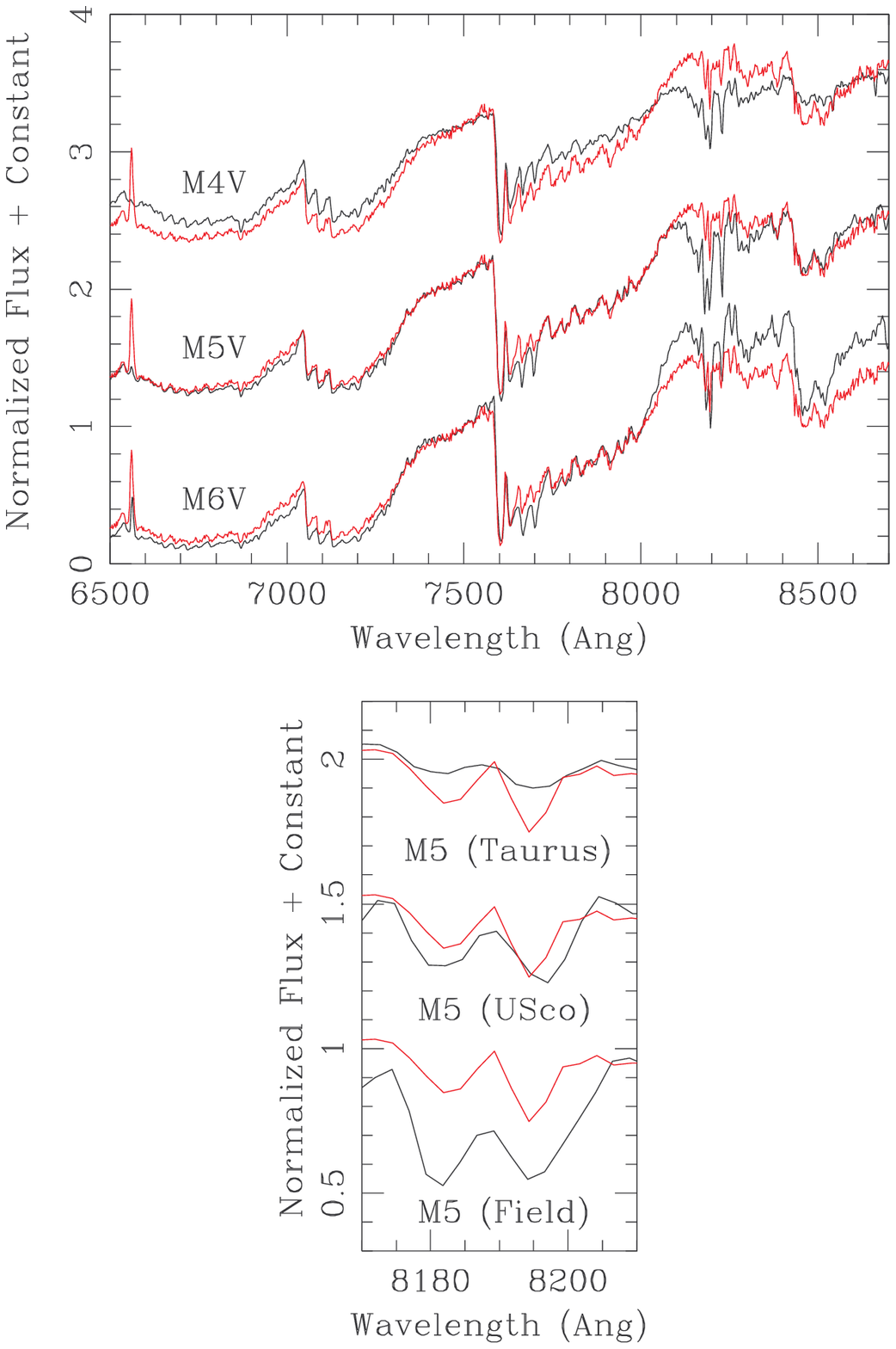}
\caption{The spectrum of USco1606-1935 B (red) as compared to a set of 
standard stars drawn from the field and from the young Taurus and Upper Sco 
associations. The overall continuum shape is best fit by a field standard with 
spectral type M5; the spectrum around the Na doublet at 8189 angstroms is 
better fit by an intermediate-age (5 Myr) M5 than a young (1-2 Myr) or field 
M5, suggesting that the B component is also intermediate-aged.}
\end{figure}

\subsection{Images}

In Figure 1, we show a NIRC2 wide-camera image of the field surrounding 
USco1606-1935. The A and B components are labeled, as are 6 apparent field 
stars (named F1 through F6) which we use as astrometric comparison 
stars. We found counterparts for the first three field stars in existing 
survey catalogues: F1 was detected by all four sky surveys, F2 was detected by 
DENIS, USNO-B, and SSS, and F3 was detected only by USNO-B and SSS.

In Figure 2, we show individual contour plots drawn from NIRC2 
narrow-camera images of the A and B components. These high-resolution 
images show that USco1606-1935 A is itself composed of two sources; we 
designate these two components Aa and Ab. We do not possess any direct 
diagnostic information to determine if Aa and Ab are physically 
associated, but there are only two other bright sources in the field of 
view. If the source count is representative of the surface density of 
bright ($K<15$) sources along the line of sight, the probability of 
finding an unbound bright source within $<$100 mas of the A component is 
only $\sim10^{-5}$. Thus, we consider Aa and Ab to comprise a physically 
bound binary system.

\subsection{Photometry}

Photometric data are generally sufficient to reject most nonmember 
interlopers because association members follow a bright, well-defined 
cluster sequence in color-magnitude diagrams and most field stars will 
fall below or bluer than the association sequence. In Table 2, we 
summarize the observed and archival photometry for each source in the 
NIRC2 wide-camera images. In Figure 3, we show three color-magnitude 
diagrams ($K$ versus $J-K$, $H-K$, and $I2-K$) for our observed sources 
and for all spectroscopically-confirmed members of Upper Sco (as 
summarized in Kraus \& Hillenbrand 2007).

The colors and magnitudes for USco1606-1935 B are consistent with the 
known members of Upper Sco, which supports the assertion that it is an 
association member. B is located marginally above and redward of the mean 
cluster sequence in the ($K$,$J-K$) and ($K$,$H-K$) diagrams; if this 
result is genuine and not a consequence of the photometric uncertainties, 
it could be a consequence of differential reddening, a $K$ band excess 
associated with a hot disk, or the presence of an unresolved tight binary 
companion. However, B does not appear to be as red in DENIS data 
($J-K=0.98$), which suggests that the 2MASS result may not be genuine.

The three sources for which we have colors (F1, F2, and F3) all sit below the 
Upper Sco member sequence in the ($K$,$I2-K$) color-magnitude diagram. 
Some USco members also fall marginally blueward of the association 
sequence in ($K$,$I2-K$); we can find no correlation with location, 
multiplicity, or other systematic factors, so this feature may be a 
result of intrinsic variability between the epochs of $K$ and $I2$. This 
result suggests that the ($K$,$I2-K$) CMD is not sufficient for ruling 
out the membership of F1. However, F1 also sits at the extreme blueward 
edge of the association sequence in ($K$,$J-K$) and is clearly distinct 
from the association sequence in ($K$,$H-K$). We therefore judge that all 
three sources are unassociated field star interlopers.

We do not possess sufficient information to determine whether these three 
stars are field dwarfs in the Milky Way disk or background giants in the 
Milky Way bulge; the unknown nature of these sources could complicate 
future efforts to calculate absolute proper motions because comparison to 
nonmoving background giants is the best way to establish a nonmoving 
astrometric frame of reference. As we will show in Section 3.3, F1 
possesses a small total proper motion ($<$10 mas yr$^{-1}$), so it may be 
a distant background star. Its 2MASS colors ($J-H=0.65$, $H-K=0.12$) place 
it on the giant sequence in a color-color diagram, but reddened early-type 
stars with spectral type $<$M0 can also reproduce these colors.

We are unable to measure colors for the stars F4, F5, and F6 because they 
were detected only in our LGSAO observations. However, visual inspection 
of the digitized POSS plates via Aladdin (Bonnarel et al. 2000) found 
possible $R$ band counterparts to F4 and F6 that were not identified by 
USNO-B. If these detections are genuine and these two sources fall near 
the USNO-B survey limit ($R\sim20-21$), their colors ($R-K\sim3-4$ or 
$I2-K\sim2-3$) are too blue to be consistent with association membership.

\subsection{Astrometry}

The standard method for confirming physical association of candidate 
binary companions is to test for common proper motion. This test is not 
as useful for young stars in associations because other (gravitationally 
unbound) association members have similar proper motions to within 
$\la$2-3 mas yr$^{-1}$. However, proper motion analysis can still be 
used to eliminate nearby late-type field stars and background giants that 
coincidentally fall along the association color-magnitude sequence but 
possess distinct kinematics.

In Table 2, we summarize the relative astrometry for the three system 
components and for the field stars F1-F6 as measured with our LGSAO 
observations and archival data from 2MASS and DENIS. All offsets are given 
with respect to the photocenter of the unresolved Aab system; Aa and Ab 
have similar fluxes and do not appear to be variable in any of these 
measurements (Section 2.3), so this zero point should be consistent 
between different epochs. We evaluated the possibility of including 
astrometric data from older photographic surveys like USNO-B and SSS, but 
rejected this idea after finding that the two surveys reported very large 
(up to 1\arcsec) differences in the separation of the A-B system from 
digitization of the same photographic plates. We calculated relative 
proper motions in each dimension by averaging the four first-epoch values 
(2MASS and DENIS; Table 2), then comparing the result to our second-epoch 
observation obtained with LGSAO. We did not attempt a least-squares fit 
because the 2MASS values are coeval and the DENIS results were measured 
only 35 days after the 2MASS results.

In Figure 4, we plot the relative astrometry between A and B and between A 
and F1 as measured by 2MASS, DENIS, and our LGSAO survey. We also show the 
expected relative motion curve if B or F1 are nonmoving background stars and 
A moves with the mean proper motion and parallax of Upper Sco, 
($\mu_{\alpha}$,$\mu_{\delta}$)=(-9.3,-20.2) mas yr$^{-1}$ and $\pi$=7 mas 
(de Zeeuw et al. 1999; Kraus \& Hillenbrand 2007). The total relative motion 
of B over the 6.8 year observation interval is (+24$\pm$25,-40$\pm$25) mas; 
the corresponding relative proper motion is (+3.5$\pm$3.7,-5.9$\pm$3.7) mas 
yr$^{-1}$, which is consistent with comovement to within $<$2$\sigma$. This 
result is inconsistent with the hypothesis that B is a nonmoving background 
star at the $8\sigma$ level. 

The relative motion of F1 is (+17$\pm$25,+105$\pm$25) mas or 
(+2.5$\pm$3.7,+15.4$\pm$3.7) mas yr$^{-1}$, which is inconsistent with 
comovement at the 4$\sigma$ level. The absolute proper motion of F1, 
assuming A moves with the mean proper motion of Upper Sco, is 
(-7$\pm$4,-5$\pm$4) mas yr$^{-1}$, which is consistent with nonmovement 
to within $<$2$\sigma$. The implication is that F1 is probably a distant 
background star, either a giant or a reddened early-type star.

\subsection{Spectroscopy}

The least ambiguous method for identifying young stars is to observe 
spectroscopic signatures of youth like lithium or various 
gravity-sensitive features. Spectroscopic confirmation is not strictly 
necessary in the case of USco1606-1935 since we confirmed common proper 
motion for the A-B system, but a spectral type is also useful in 
constraining the physical properties of the secondary, so we decided to 
obtain an optical spectrum.

In the top panel of Figure 5, we plot our spectrum for B in comparison to 
three standard field dwarfs with spectral types of M4V-M6V. We qualitatively 
find that the standard star which produces the best fit is GJ 866 (M5V). The 
M4V and M6V standards do not adequately fit either the overall continuum shape 
or the depths of the TiO features at 8000 and 8500 angstroms, so the 
corresponding uncertainty in the spectral type is $\la$0.5 subclasses.

In the bottom panel of Figure 5, we plot a restricted range of the spectrum 
(8170-8210 angstroms) centered on the Na-8189 absorption doublet. The depth 
of the doublet is sensitive to surface gravity (e.g. Slesnick et al. 2006a, 
2006b); high-gravity dwarfs possess very deep absorption lines, while 
low-gravity giants show almost no absorption. We also plot standard stars of 
identical spectral type (M5) spanning a range of ages. The depth of the B 
component's Na 8189 doublet appears to be consistent with the depth for a 
member of USco (5 Myr), deeper than that of a Taurus member (1-2 Myr), and 
shallower than that of a field star, which confirms that the B component is 
a pre-main sequence member of Upper Sco.

We have quantified our analysis by calculating the spectral indices 
TiO-7140, TiO-8465, and Na-8189, which measure the depth of key temperature- 
and gravity-sensitive features (Slesnick et al. 2006a). We find that 
$TiO_{7140}=2.28$, $TiO_{8465}=1.23$, and $Na_{8189}=0.92$; all three 
indices are consistent with our assessment that B is a young M5 star which 
has not yet contracted to the zero-age main sequence.

\subsection{Stellar and Binary Properties}

\begin{deluxetable}{lcc}
\tabletypesize{\scriptsize}
\tablewidth{0pt}
\tablecaption{Binary Properties}
\tablehead{\colhead{Property} & \colhead{Aa-Ab} & \colhead{A-B}
}
\startdata
Measured\\
Sep (mas)&53.2$\pm$1.0&10874$\pm$5\\
PA (deg)&38.7$\pm$1.0&226.45$\pm$0.03\\
$\Delta$$K$ (mag)&0.17$\pm$0.05&0.70$\pm$0.05\\
$a_{proj}$ (AU)&7.7$\pm$1.2&1600$\pm$200\\
Inferred\\
$q$&0.88$\pm$0.05&0.53$\pm$0.08\\
$SpT_{Prim}$&M5$\pm$0.5&M5+M5.2($\pm$0.5)\\
$SpT_{Sec}$&M5.2$\pm$0.5&M5$\pm$0.5\\
$M_{Prim}$&0.14$\pm$0.02&0.26$\pm$0.04\\
$M_{Sec}$&0.12$\pm$0.02&0.14$\pm$0.02\\
\enddata
\tablecomments{The center of mass for the Aa-Ab pair is unknown, so we calculate 
all A-B separations with respect to the $K$ band photocenter.}
\end{deluxetable}

In Table 3, we list the inferred stellar and binary properties for the 
Aa-Ab and A-B systems, which we estimate using the methods described in 
Kraus \& Hillenbrand (2007). This procedure calculates component masses by 
combining the 5 Myr isochrone of Baraffe et al. (1998) and the M dwarf 
temperature scale of Luhman et al. (2003) to directly convert observed 
spectral types to masses. Relative properties (mass ratios $q$ and 
relative spectral types) are calculated by combining the Baraffe 
isochrones and Luhman temperature scale with the empirical NIR colors of 
Bessell \& Brett (1998) and the K-band bolometric corrections of Leggett 
et al. (1998) to estimate $q$ and $\Delta$$SpT$ from the observed flux 
ratio $\Delta$$K$. 

We have adopted the previously-measured spectral type for A (M5; 
Preibisch et al. 2002) as the type for component Aa, but the inferred 
spectral type for Ab is only 0.2 subclasses later, so this assumption 
should be robust to within the uncertainties ($\sim$0.5 subclasses). The 
projected spatial separations are calculated for the mean distance of 
Upper Sco, 145$\pm$2 pc (de Zeeuw et al. 1999). If the total radial 
depth of Upper Sco is equal to its angular extent ($\sim$15$^o$ or 
$\sim$40 pc), then the unknown depth of USco1606-1935 within Upper Sco 
implies an uncertainty in the projected spatial separation of $\pm$15\%. 
The systematic uncertainty due to the uncertainty in the mean distance of 
Upper Sco is negligible ($\la$2\%).

\section{Is USco1606-1935 AB a Binary System?}

\begin{figure*}
\epsscale{1.00}
\plotone{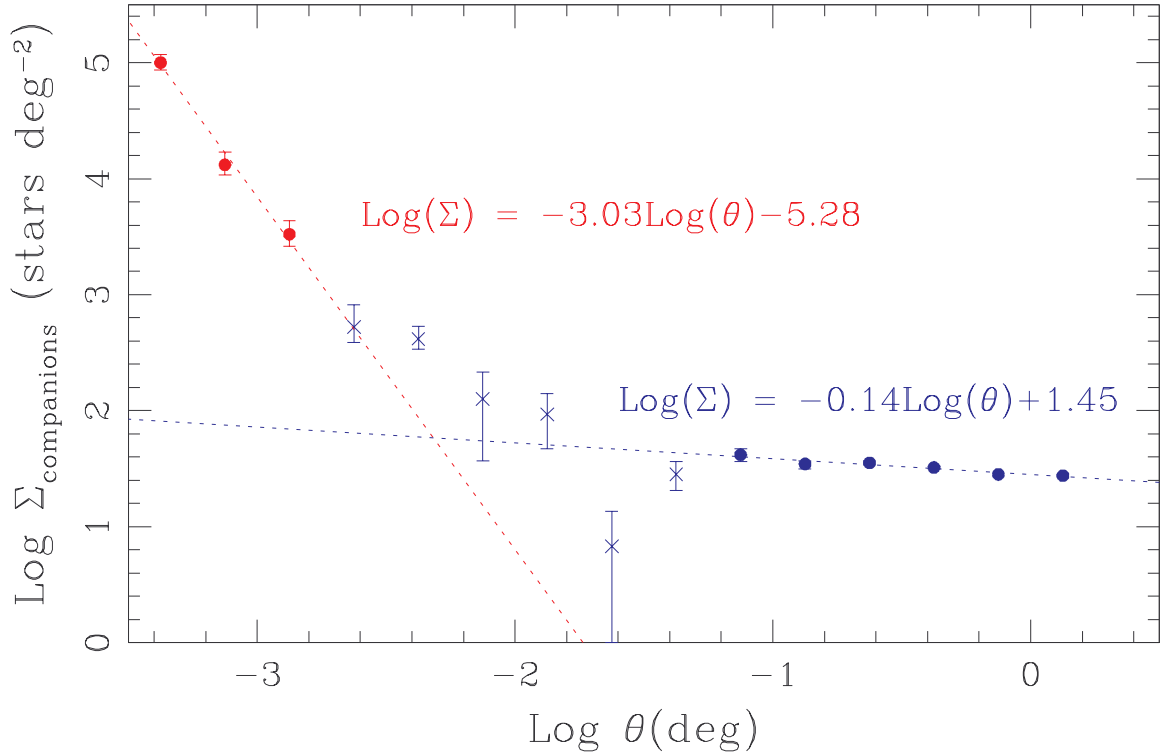}
\caption{The surface density of companions as a function of separation for 
young stars and brown dwarfs in Upper Sco. Red symbols denote results from our 
wide-binary survey using 2MASS (Kraus \& Hillenbrand 2007) and blue symbols 
denote data for all spectroscopically-confirmed members in two fields surveyed 
by Preibisch et al. (2002). The data appear to be well-fit by two power 
laws (dashed lines) which most likely correspond to gravitationally bound 
binaries and unbound clusters of stars that have not yet completely 
dispersed from their formation environments. The data points which were 
used to fit these power laws are denoted with circles; other points are 
denoted with crosses.
}
\end{figure*}

The unambiguous identification of pre-main sequence binaries is 
complicated by the difficulty of distinguishing gravitationally bound 
binary pairs from coeval, comoving association members which are aligned 
in projection. Most traditional methods used to confirm field binary 
companions do not work in the case of young binaries in clusters and 
associations because all association members share common distances and 
kinematics (to within current observational uncertainties), so the only 
remaining option is to assess the probability of chance alignment. We 
address this challenge by quantifying the clustering of PMS stars via 
calculation of the two-point correlation function (TPCF) across a wide 
range of angular scales (1\arcsec\, to $>$1 degree).  This type of 
analysis has been attempted in the past (e.g. Gomez et al. 1993 for 
Taurus; Simon 1997 for Ophiuchus, Taurus, and the Trapezium), but these 
studies were conducted using samples that were significantly incomplete 
relative to today.

The TPCF, $w(\theta)$, is defined as the number of excess pairs of 
objects with a given separation $\theta$ over the expected number for a 
random distribution (Peebles 1980). The TPCF is linearly proportional to 
the surface density of companions per star, 
$\Sigma(\theta)=(N_*/A)[1+w(\theta)]$, where $A$ is the survey area and 
$N_*$ is the total number of stars. However, it is often easier to 
evaluate the TPCF via a Monte Carlo-based definition, 
$w(\theta)=N_p(\theta)/N_r(\theta)-1$, where $N_p(\theta)$ is the number 
of pairs in the survey area with separations in a bin centered on 
$\theta$ and $N_r(\theta)$ is the expected number of pairs for a random 
distribution of objects over the same area (Hewett 1982). The advantage 
of this method is that it does not require edge corrections, unlike 
direct measurement of $\Sigma(\theta)$. We adopted this method due to its 
ease of implementation, but we report our subsequent results in terms of 
$\Sigma(\theta)$ since it is a more intuitive quantity.

The current census of Upper Sco members across the full association is 
very incomplete, so we implemented our analysis for intermediate and large 
separations ($\theta>6.4\arcsec$) using only members located in two 
heavily-studied fields originally observed by Preibisch et al. (2001, 
2002; the 2df-East and 2df-West fields). The census of members in these 
fields may not be complete, but we expect that it is the least incomplete. 
The census of companions at smaller separations (1.5-6.4\arcsec) has 
been uniformly studied for all spectroscopically-confirmed members (Kraus 
\& Hillenbrand 2007), so we have maximized the sample size in this 
separation range by considering the immediate area around all known 
members, not just those within the Preibisch fields. Our survey was only 
complete for mass ratios $q>$0.25, so we do not include companions with 
mass ratios $q<0.25$.

These choices might lead to systematic biases if the Preibisch fields are 
still significantly incomplete or if the frequency and properties of 
binary systems show intra-association variations, but any such 
incompleteness would probably change the result by no more than a factor 
of 2-3. As we will subsequently show, $\Sigma(\theta)$ varies by 4 orders 
of magnitude across the full range of $\theta$. The well-established mass 
dependence of multiplicity should not affect our results since the mass 
function for the Preibisch fields is similar to that seen for the rest of 
the association.

In Figure 6, we plot $\Sigma(\theta)$ for Upper Sco, spanning the 
separation range $-3.5<log(\theta)<0.25$ (1.14\arcsec\, to 1.78 deg ). We 
have fit this relation with two power laws, one which dominates at small 
separations ($\la$15-30\arcsec) and one at larger separations. We 
interpret the two segments, following Simon (1997), to be the result of 
gravitationally-bound binarity and gravitationally unbound 
intra-association clustering, respectively. We fit the binary power law to 
the three lowest-separation bins ($log(\theta)<-2.75$) because this is the 
separation range over which we possess uniform multiplicity data. The 
cluster power law was fit to the six highest-separation bins 
($log(\theta)>-1.25$) because those bins have the smallest uncertainties. 
Bins corresponding to intermediate separations seem to follow the two 
power laws.

We found that the slope of the cluster power law (-0.14$\pm$0.02) is very 
close to zero, which implies that there is very little clustering on 
scales of $\la$1 deg. This result is not unexpected for intermediate-age 
associations like Upper Sco; given the typical intra-association velocity 
dispersion ($\sim$1 km s$^{-1}$) and the age (5 Myr), most association 
members have dispersed $\sim$5 pc (2 deg) relative to their formation 
point, averaging out structure on smaller spatial scales. Simon (1997) 
found that the slopes for Taurus, Ophiuchus, and the ONC are steeper, 
suggesting that more structure is present on these small scales at young 
ages ($\sim$1-2 Myr). The slope of the binary power law (-3.03$\pm$0.24) 
is much steeper than the cluster regime. The separation range represented 
is much larger than the peak of the binary separation distribution 
($\sim$30 AU for field solar-mass stars; Duquennoy \& Mayor 1991), so the 
steep negative slope corresponds to the large-separation tail of the 
separation distribution function. The two power laws seem to cross at 
separations of $\sim$15-30\arcsec\, ($a_{proj}\sim2500-5000$ AU), though 
this result depends on the sample completeness in the binary and cluster 
regimes. We interpret this to be the maximum separation range at which 
binaries can be identified.
 
If we extrapolate the cluster power law into the separation regime of the 
binary power law, we find that the expected surface density of unbound 
coincidentally-aligned companions is $\sim$60 deg$^{-2}$. Given this 
surface density, there should be $\sim$1 chance alignment within 
15\arcsec\, among the 366 spectroscopically confirmed members of Upper 
Sco. Among the 173 known late-type stars and brown dwarfs (SpT$\ge$M4) for 
which this separation range is unusually wide, the expected number of 
chance alignments with any other member is 0.5. If the mass function of 
known members is similar to the total mass function, approximately half 
($\sim$0.25 chance alignments) are expected to occur with another low-mass 
member. Therefore, we expect $\sim$0.25 chance alignments which might be 
mistaken for a low-mass binary pair.

The probability that one or more such chance alignments actually exists 
for a known low-mass USco member is 25\% (based on Poisson statistics), 
which suggests that the nature of a single candidate wide pair like 
USco1606-1935 AB can not be unambiguously determined. If any more pairs 
can be confirmed, then they would represent a statistically significant 
excess. The corresponding probability of finding 2 chance alignments of 
low-mass members is only 2\%. As we have described in our survey of wide 
multiplicity with 2MASS (Kraus \& Hillenbrand 2007), we have identified at 
least three additional candidate ultrawide systems in Upper Sco, so 
spectroscopic and astrometric followup of these candidate systems is a 
high priority.

\section{Summary}

We have presented photometric, astrometric, and spectroscopic 
observations of USco1606-1935, a candidate ultrawide ($\sim$1600 AU), 
low-mass ($M_{tot}\sim$0.4 $M_{\sun}$) hierarchical triple system in the 
nearby OB association Upper Scorpius. We conclude that the ultrawide B 
component is a young, comoving member of the association, and show that 
the primary is itself a close binary system.

If the Aab and B components are gravitationally bound, the system would 
join the growing class of young multiple systems which have unusually 
wide separations as compared to field systems of similar mass. However, 
we demonstrate that binarity can not be assumed purely on probabilistic 
grounds. Analysis of the association's two-point correlation function 
shows that there is a significant probability (25\%) that at least one 
pair of low-mass association members will be separated by $\la$15\arcsec, 
so analysis of the wide binary population requires a systematic search 
for all wide binaries. The detection of another pair of low-mass members 
within 15\arcsec\, would represent an excess at the 98\% confidence 
level. In principle, binarity could also be demonstrated by measuring 
common proper motion with precision higher than the internal velocity 
scatter of the association; given the astrometric precision currently 
attainable with LGSAO data ($\la$1 mas), the test could be feasible 
within $\la$5 years.

\acknowledgements

The authors thank C. Slesnick for providing guidance in the analysis of young 
stellar spectra, P. Cameron for sharing his NIRC2 astrometric calibration 
results prior to publication, and the anonymous referee for returning a helpful 
and very prompt review. The authors also wish to thank the observatory staff, 
and particularly the Keck LGSAO team, for their tireless efforts in 
commissioning this valuable addition to the observatory. Finally, we recognize 
and acknowledge the very significant cultural role and reverence that the 
summit of Mauna Kea has always had within the indigenous Hawaiian community.  
We are most fortunate to have the opportunity to conduct observations from this 
mountain.

This work makes use of data products from the Two Micron All-Sky Survey, which is 
a joint project of the University of Massachusetts and the Infrared Processing 
and Analysis Center/California Institute of Technology, funded by the National 
Aeronautics and Space Administration and the National Science Foundation. This 
work also makes use of data products from the DENIS project, which has been 
partly funded by the SCIENCE and the HCM plans of the European Commission under 
grants CT920791 and CT940627. It is supported by INSU, MEN and CNRS in France, by 
the State of Baden-Württemberg in Germany, by DGICYT in Spain, by CNR in Italy, 
by FFwFBWF in Austria, by FAPESP in Brazil, by OTKA grants F-4239 and F-013990 in 
Hungary, and by the ESO C\&EE grant A-04-046. Finally, our research has made use 
of the USNOFS Image and Catalogue Archive operated by the United States Naval 
Observatory, Flagstaff Station (http://www.nofs.navy.mil/data/fchpix/).

\end{document}